\documentstyle[epsf]{elsart}
\newcommand{\vecc}[1]{\mbox{\boldmath $#1$}}

\begin{document}

\begin{frontmatter}

\title{Emission of Two Hard Photons in
Large--Angle Bhabha Scattering }

\author[LTPH]{A.B.~Arbuzov},
\author[NSK]{V.A.~Astakhov},
\author[LTPH]{E.A.~Kuraev},
\author[PTI]{N.P.~Merenkov},
\author[Parma]{L.~Trentadue},
\author[LVTA]{E.V.~Zemlyanaya}
\address[LTPH]{Bogoliubov Laboratory of Theoretical Physics, JINR,
Dubna, 141980, Russia}
\address[NSK]{Budker Institute for Nuclear Physics,
Prospect Nauki 11, Novosibirsk, 630090, Russia}
\address[PTI]{Institute of Physics and Technology,
Kharkov, 310108, Ukraine}
\address[Parma]{Dipartimento di Fisica, Universit\'a di Parma, 43100 Parma;\\
INFN Sezione di Milano, Milano, Italy}
\address[LVTA]{Laboratory of Computing Techniques
and Automation, JINR, Dubna, 141980, Russia}

\begin{abstract}
A closed expression for the differential cross section
of the large--angle Bhabha $e^+ e^-$ scattering which explicitly
takes into account the leading and next--to--leading contributions
due to the emission of two hard photons is presented.
Both collinear and semi--collinear kinematical regions are considered.
The results are illustrated by numerical calculations.
\end{abstract}

\begin{keyword}
Bhabha scattering, double bremsstrahlung, large angles, high energy
\end{keyword}

\end{frontmatter}

\vspace{.5cm}

PACS~ 12.20.-m Quantum electrodynamics, 12.20.Ds Specific calculations

\vspace{1.5cm}

\noindent
{\small Corresponding author: A.B. Arbuzov, \\
E-mail: arbuzov@thsun1.jinr.dubna.su \\
FAX: (+007) 096 21 65084 \\
Telephone: (+007) 096 21 63047  }

\thispagestyle{empty}
\setcounter{page}{0}

\newpage
\setcounter{page}{1}

\section{Introduction}

The large--angle Bhabha process is well suited for the determination
of the luminosity ${\cal{L}}$ at $e^+e^-$ colliders of the
intermediate energy range $\sqrt{s}=2 \varepsilon \sim 1 \mbox{GeV}$
\cite{vepp2m,dafne}. Small scattering angle kinematics of Bhabha scattering
is used for high--energy colliders such as LEP~I \cite{sabs}.
As far as $0.1 \%$ accuracy is desirable
in the determination of ${\cal L}$, the corresponding requirement
\begin{eqnarray}
\bigg| \frac{\delta\sigma}{\sigma}  \bigg| \leq 10^{-3}
\end{eqnarray}
on the Bhabha cross section theoretical description appears.
The quantity $\Delta\sigma$ is an unknown uncertainty in the cross section
due to higher order radiative corrections.
A great attention was paid to this process during the
last decades (see review~\cite{labsr} and references therein).
The Born cross section with weak interactions taken
into account and the first order QED radiative corrections to it
were studied in detail~\cite{born}.
Both contributions, the one enhanced by
{\em the large logarithmic multiplier\/} $L=\ln(s/m^2)\ $
(where $s=(p_++p_-)^2=4\varepsilon^2$ is the total
center--of--mass (CM) energy squared, $m$ is the electron mass),
and the one without $L$ are to
be kept in the limits (1): $\alpha L/\pi$, $\alpha/\pi$.
As for the corrections in the second order of the perturbation theory,
they are necessary in the leading and next--to--leading approximations
and take the following orders, respectively:
\begin{eqnarray}
\biggl(\frac{\alpha}{\pi}\biggr)^2L^2, \quad
\biggl(\frac{\alpha}{\pi}\biggr)^2L.
\end{eqnarray}
The total two--loop ($\sim (\alpha /\pi)^2$) correction could be
constructed from:
1) the two--loop corrections arising from the emission of
two virtual photons;
2) the one--loop corrections to a single real (soft and hard)
photon emission;
3) the ones arising from the emission of two real photons;
4) the virtual and real $e^+e^-$ pair production~\cite{pairs}.
As for the corrections in the third order of perturbation theory,
only the leading ones proportional to $(\alpha L /\pi)^3$
are to be taken into account.

In this paper we consider the emission of two real hard photons:
\begin{equation}
e^+(p_+)+e^-(p_-)\rightarrow e^+(q_+)+e^-(q_-)+\gamma(k_1)+\gamma(k_2).
\label{proc}
\end{equation}
The relevant contribution to the {\em experimental\/} cross section has the
following form
\begin{equation}
\sigma_{\mathrm{exp}}=\int \d\sigma \; \Theta_+\Theta_-,
\end{equation}
where $\Theta_+$ and $\Theta_-$ are the experimental restrictions
providing the simultaneous detection of both the scattered electron
and positron. First, this means that their energy fractions
should be larger than a certain (small) quantity
$\varepsilon_{\mathrm{th}}/\varepsilon$, $\varepsilon_{\mathrm{th}}$
is the energy threshold of the detectors.
The second condition restricts their
angles with respect to the beam axes. They should be larger
than a certain finite value $\psi_0$ ($\ \psi_0\sim 35^{\circ}$
in the experimental conditions accepted in \cite{vepp2m}):
\begin{equation}
\pi - \psi_0 > \theta_{-}, \, \theta_{+}> \psi_0, \qquad
\theta_{\pm}=\widehat{\vecc{q}_{\pm}\vecc{p}}_{-}\, ,
\end{equation}
where $\theta_{\pm}$ are the polar angles of the scattered
leptons with respect to the beam axes ($\vecc{p}_-$).
We accept the condition on the energy threshold
of the charged--particle registration
$q_{\pm}^0 > \varepsilon_{\mathrm{th}}$.
Both photons are assumed to be hard. Their minimal energy
\begin{eqnarray}
\omega_{\mathrm{min}}=\Delta\varepsilon, \qquad \Delta \ll 1,
\end{eqnarray}
could be considered as the threshold of the photon registration.

The main $(\sim (\alpha L/\pi)^2$) contribution to the total
cross section (5) comes from the collinear region:
when both the emitted photons move within
narrow cones along the charged particle momenta (they may
go along the same particle). So
we will distinguish 16 kinematical regions:
\begin{eqnarray}
&& \widehat{\vecc{a} \vecc{k}}_1\ \ \mbox{and}\ \
\widehat{\vecc{a} \vecc{k}}_2 < \theta_0,
\qquad
\widehat{\vecc{a} \vecc{k}}_1\ \ \mbox{and}\ \
\widehat{\vecc{b} \vecc{k}}_2 < \theta_0,
\nonumber \\ \label{eq6}
&& \frac{m}{\varepsilon} \ll \theta_0 \ll 1, \qquad
 a \ne b, \quad a,\, b=p_-, p_+, q_-, q_+ \, .
\end{eqnarray}
The matrix element module square summed over spin states in the regions
(\ref{eq6}) is of the form of the Born matrix element multiplied by the
so--called collinear factors. The contribution
to the cross section of each region has also the form of $2\to 2$
Bhabha cross sections in the Born approximation multiplied by factors of
the form
\begin{equation}
\d\sigma_i^{\mathrm{coll}}=\d\sigma_{0i} \biggl[ a_i(x_j,y_j)
\ln^2\biggl(\frac{\varepsilon^2\theta_0^2}{m^2}\biggr) + b_i(x_j,y_j)
\ln\biggl(\frac{\varepsilon^2\theta_0^2}{m^2}\biggr) \biggr],
\end{equation}
where $x_j=\omega_j/\varepsilon$, $y_1=q_{-}^0/\varepsilon$,
$y_2=q_{+}^0/\varepsilon$ are the energy fractions of the photons
and of the scattered electron and positron.
The dependence on the auxiliary parameter $\theta_0$ will be cancelled
in the sum of the contributions of the collinear and semi--collinear regions.
The last region corresponds to the kinematics, when only one photon
is emitted inside the narrow cone $\theta_1 < \theta_0$ along one of
the charged particle momenta. And the second photon is emitted outside
any cone of that sort along charged particles ($\theta_2 > \theta_0$):
\begin{equation} \label{scol}
\d \sigma_i^{\mathrm{sc}}=\frac{\alpha}{\pi}
\ln\biggl(\frac{4\varepsilon^2}{m^2}\biggr)
\d \sigma_{0i}^{\gamma}(k_2),
\end{equation}
where $\d \sigma_{0i}^{\gamma}$ has the known form of the single hard
bremsstrahlung cross section in the Born approximation \cite{brem}.

Below we show explicitly that the result of the integration over the
single hard photon emission in eq. (\ref{scol}) in the kinematical
region $\theta_2^i > \theta_0$ ($\theta_2^i$ is the emission angle of
the second hard photon with respect to the direction of one of the
four charged particles) has the following form
\begin{equation}
\int \d \sigma_{0i}^{\gamma}(k_2)=-2\ln\biggl(\frac{\theta_0^2}{4}\biggr)
a_i(x,y) \d\sigma_0^i + \d\tilde{\sigma}^i.
\end{equation}

The collinear factors in the double bremsstrahlung process
were first considered in papers of the CALCUL collaboration~\cite{calcul}.
Unfortunately they have a rather complicated form
which is less convenient for further analytical integration
in comparison with the expressions given below.
The method of calculation of the collinear factors may
be considered as a generalization of the quasi--real electron
method~\cite{quasi} to the case of multiple bremsstrahlung.
Another generalization is required for the calculations of
the cross section of process $e^+e^- \to 2e^+2e^-$ \cite{pairs}.

It is interesting that the collinear factors for the
kinematical region of the two hard photon emission along
the projectile and the scattered electron are found the
same as for the electron--proton scattering process considered
by one of us (N.P.M.) in paper~\cite{npm}.

There are 40 Feynman diagrams of the tree type which describe the double
bremsstrahlung process in $e^+e^-$ collisions. The differential
cross section in terms of helicity amplitudes was computed
about ten years ago~\cite{calcul,kurper}.
It has a very complicated form. We note that the contribution from
the kinematical region in which
the angles (in the CM system) between any two final
particles are large compared with $m/\varepsilon$ is of the order
\begin{equation}
\frac{\alpha^2 r_0^2 m^2}{\pi^2\varepsilon^2} \sim 10^{-36} \mbox{cm}^2,
\end{equation}
($r_0$ is the classical electron radius).
So, the corresponding events will possess poor statistics at
the colliders with the luminosity
${\cal L} \sim 10^{31}\, - \, 10^{32} \mbox{cm}^{-2}\mbox{s}^{-1}$.
More probable are the cases of double bremsstrahlung imitating
the processes $e^+e^- \to e^+e^-$ or $e^+e^- \to e^+e^-\gamma$,
which corresponds to the emission of one or two photons along
charged--particle momenta.

\section{Kinematics in the collinear region}

It is convenient to introduce, in the collinear region, new variables
and transform the phase volume of the final state in the following
way (from now on we will work in the CM system):
\begin{eqnarray}
&& \int\! \d \Gamma = \int\! \frac{\d^3q_-\d^3q_+\d^3k_1\d^3k_2}
{16q^0_{-}q^0_{+}\omega_1\omega_2(2\pi)^8}
\delta^{(4)}(\eta_1p_- + \eta_2p_+ - \lambda_1q_-  - \lambda_2q_+)
\nonumber \\ && \qquad  \label{z0}
=\frac{m^4\pi^2}{4(2\pi)^6}\int\limits_{\Delta}^{1}\! \d x_1
\int\limits_{\Delta}^{1}\! \d x_2\; x_1 x_2
\int\limits_{0}^{2\pi}\! \frac{\d\phi}{2\pi}
\int\limits_{0}^{z_0}\! \d z_1 \int\limits_{0}^{z_0}\! \d z_2
\int\! \d \Gamma_q,
\\ \nonumber &&
\int\! \d \Gamma_q=\int \frac{\d^3q_-\d^3q_+}
{4q^0_{-}q^0_{+}(2\pi)^2}
\delta^{(4)}(\eta_1p_- + \eta_2p_+ - \lambda_1q_-  - \lambda_2q_+),
\\ \nonumber &&
z_{1,2}=\biggl( \frac{\theta_{1,2}\varepsilon}{m} \biggr)^2,
\quad \phi= \widehat { \vecc{k}_{1\bot} \vecc{k}}_{2\bot},
\quad x_i=\frac{\omega_i}{\varepsilon}, \quad
z_0 = \biggl( \frac{\theta_{0}\varepsilon}{m} \biggr)^2 \gg 1,
\quad \Delta=\frac{\omega_{\mathrm{min}}}{\varepsilon}\, ,
\end{eqnarray}
where $\theta_i$ $(i=1,2)$ is the polar angle of the $i$--photon
emission with respect to the momentum of the charged
particle that emits the photon; $\eta_{\pm}$ and
$\lambda_{\pm}$ depend on the specific emission kinematics,
they are given in Table 1.

\vspace{.5cm}
\begin{table}
\caption{$\eta_i$ and $\lambda_i$
for different collinear kinematics.}
\begin{tabular}{|c|c|c|c|c|c|c|c|c|c|c|}
\hline
~~ & $p_-p_-$ & $q_-q_-$ & $p_+p_+$ & $q_+q_+$ & $p_-p_+$ &
$q_-q_+$ & $p_-q_-$ & $p_+q_+$ & $p_-q_+$ & $p_+q_-$  \\ \hline
$\eta_1$ & $y$ & $1$ & $1$ & $1$ & $1-x_1$ & $1$ & $1-x_1$ & $1$ &
$1-x_1$ & $1$  \\ \hline
$\eta_2$ & $1$ & $1$ & $y$ & $1$ & $1-x_2$ & $1$ & $1$ & $1-x_1$ &
$1$ & $1-x_1$  \\ \hline
$\lambda_1$ & $1$ & $\frac{1}{y}$ & $1$ & $1$ & $1$ & $\frac{1}{1-x_1}$
& $1+\frac{x_2}{y_1}$ & $1$ & $1$ & $1+\frac{x_2}{y_1}$  \\ \hline
$\lambda_2$ & $1$ & $1$ & $1$ & $\frac{1}{y}$ & $1$ & $\frac{1}{1-x_2}$
& $1$ & $1+\frac{x_2}{y_2}$ & $1+\frac{x_2}{y_2}$ & $1$  \\ \hline
\end{tabular}
\end{table}

\vspace{.5cm}

The columns of the Table correspond to a certain choice of
the kinematics in the following way: $p_-p_-$ means the emission
of both the photons along the projectile electron,
$p_+q_-$ means that the first of the photons goes along the
projectile positron; the second, along the scattered
electron, and so on. The contributions from 6 remaining kinematical
regions (when the photons in the last 6 columns are interchanged)
could be found by the simple substitution $x_1 \leftrightarrow x_2$.
We will use the momentum conservation law
\begin{equation}
\eta_1p_- + \eta_2p_+ = \lambda_1q_- +  \lambda_2q_+ \, ,
\label{conser}
\end{equation}
and the following relations coming from it:
\begin{eqnarray} \label{eq:14}
&& \eta_1+\eta_2=\lambda_1y_1 + \lambda_2y_2, \qquad
\lambda_1y_1\sin\theta_-=\lambda_2y_2\sin\theta_+,
\qquad y_{1,2}=\frac{q_{1,2}^0}{\varepsilon}\, ,
\nonumber \\  && \label{cos}
\lambda_2 y_2=\frac{\eta_1^2+\eta_2^2+(\eta_2^2-\eta_1^2)c}
{\eta_1+\eta_2+(\eta_2-\eta_1)c}\, .
\end{eqnarray}

Each of 16 contributions to the cross section of process (\ref{proc})
can be expressed in terms of the corresponding Born--like cross section
multiplied by its collinear factor:
\begin{eqnarray} \label{di}
\d \sigma_{\mathrm{coll}} &=& \frac{1}{2!}\biggl(\frac{\alpha}{2\pi}\biggr)^2
\frac{x_1 x_2}{2} \sum_{(\eta,\lambda)} \overline{K}(\eta,\lambda)
\d \tilde{\sigma}_0(\eta,\lambda) \d x_1 \d x_2, \\ \nonumber
\d \tilde{\sigma}_0(\eta,\lambda) &=& \frac{2\alpha^2}{s}\, B(\eta,\lambda)\,
\d I(\eta,\lambda), \qquad
B(\eta,\lambda) = \biggl( \frac{\tilde{s}^2+\tilde{t}^2
+ \tilde{s}\tilde{t}}{\tilde{s}\tilde{t}} \biggr)^2,
\\ \nonumber
\d I_i(\eta,\lambda) &=& \int\!\! \frac{\d^3q_-\d^3q_+}{q_-^0q_+^0}
\delta^{(4)}(\eta_1p_-+\eta_2p_+-\lambda_1q_--\lambda_2q_+) \\ \nonumber
&=&\frac{4\pi\eta_1\eta_2\d c}{\lambda_1^2\lambda_2^2
[c(\eta_2-\eta_1)+\eta_1+\eta_2]^2}\, ,\\ \nonumber
\overline{K}(\eta,\lambda) &=& m^4 \int\limits_{0}^{z_0}\d z_1
\int\limits_{0}^{z_0}\d z_2 \int\limits_{0}^{2\pi}
\frac{\d \phi}{2\pi} {\cal K}(\eta,\lambda), \\ \nonumber
\tilde{t} &=& (\eta_1p_--\lambda_1q_-)^2
=-\tilde{s}\frac{\eta_1(1-c)}{\eta_1+\eta_2+(\eta_2-\eta_1)c}\, , \\ \nonumber
\tilde{s} &=& (\eta_1p_-+\eta_2p_+)^2
=4\varepsilon^2\eta_1\eta_2=s\eta_1\eta_2,
\quad \tilde{s}+\tilde{t}+\tilde{u}=0.
\end{eqnarray}
The sum over $(\eta,\lambda)$ means the sum over 16
collinear kinematical regions. The corresponding $(\eta,\lambda)$
could be found in Table~1. The quantities
${\cal K}_i(\eta,\lambda)$ are as follows:
\begin{eqnarray}
&& {\cal K}(p_-p_-)=\frac{2}{y}{\cal A}(A_1,A_2,A,x_1,x_2,y), \quad
{\cal K}(q_-q_-)=2y{\cal A}(B_1,B_2,B,\frac{-x_1}{y},\frac{-x_2}{y},
\frac{1}{y}), \nonumber \\ \nonumber
&& {\cal K }(p_+p_+)=\frac{2}{y}{\cal A}(C_1,C_2,C,x_1,x_2,y), \quad
{\cal K}(q_+q_+)=2y{\cal A}(D_1,D_2,D,\frac{-x_1}{y},\frac{-x_2}{y},
\frac{1}{y}), \\
&& {\cal A}(A_1,A_2,A,x_1,x_2)=-\frac{yA_2}{A^2A_1} - \frac{yA_1}{A^2A_2}
+ \frac{1+y^2}{x_1x_2A_1A_2} + \frac{r_1^3+yr_2}{AA_1x_1x_2}
\nonumber \\ && \quad \label{eq:19}
+ \frac{r_2^3+yr_1}{AA_2x_1x_2}
+ \frac{2m^2(y^2+r_1^2)}{AA_1^2x_2}
+ \frac{2m^2(y^2+r_2^2)}{AA_2^2x_1},
\end{eqnarray}
\begin{eqnarray} \label{eq:20}
&& {\cal K}(p_-p_+)=2K_1K_2,\qquad {\cal K}(p_-q_+)=-2K_1K_3,\qquad
{\cal K}(p_+q_-)=-2K_4K_5, \\ \nonumber
&& {\cal K}(q_-q_+)=2K_6K_7,\qquad {\cal K}(p_-q_-)=-2K_1K_5,\qquad
{\cal K}(p_+q_+)=-2K_4K_3, \\ \nonumber
&& K_1=\frac{g_1}{A_1x_1r_1}+\frac{2m^2}{A_1^2}, \quad
K_2=\frac{g_2}{C_2x_2r_2}+\frac{2m^2}{C_2^2}, \quad
K_3=\frac{g_4}{D_2x_2t_2}-\frac{2m^2}{D_2^2}, \\ \nonumber
&& K_4=\frac{g_1}{C_1x_1r_1}+\frac{2m^2}{C_1^2}, \quad
K_5=\frac{g_3}{B_2x_2t_1}-\frac{2m^2}{B_2^2}, \quad
K_6=\frac{g_1}{B_1x_1}-\frac{2m^2}{B_1^2}, \\ \nonumber
&& K_7=\frac{g_2}{D_2x_2}-\frac{2m^2}{D_2^2}, \qquad
r_1=1-x_1,\quad r_2=1-x_2, \\ \nonumber
&& g_1=1+r_1^2, \quad g_2=1+r_2^2,\quad g_3=y_1^2+t_1^2,
\quad  g_4=y_2^2+t_2^2, \\ \nonumber
&& t_1=y_1+x_2,\quad t_2=y_2+x_2, \quad y=1-x_1-x_2,
\end{eqnarray}
$y_1,\,y_2$ are the energy fractions of the scattered
electron and positron defined in eq.~(\ref{cos}).

Expressions (\ref{eq:20}) agree with the results of paper~\cite{calcul}
except for a simpler form of ${\cal K}(q_-q_+)$.
As for eq. (\ref{eq:19}) it has an evident advantage
in comparison to the corresponding formulae given in paper~\cite{calcul}.
Let us note that the remaining factors ${\cal K}(p,q)$ could be
obtained from the ones given in eq.~(\ref{eq:20}) using relations of
the following type:
\begin{eqnarray}
{\cal K}(p_-q_-)(x_1,x_2,A_1,B_2)={\cal K}(q_-p_-)(x_2,x_1,A_2,B_1).
\end{eqnarray}
Note also that terms of the kind $m^4/(B_2^2C_1^2)$
do not give logarithmically enhanced contributions, and
we will omit them below.
The denominators of the propagators entering into
eqs. (\ref{eq:19}), (\ref{eq:20}) are:
\begin{eqnarray}
&& A_i=(p_--k_i)^2-m^2, \qquad A=(p_--k_1-k_2)^2-m^2, \nonumber
\\ \label{abcd}
&& B_i=(q_-+k_i)^2-m^2, \qquad B=(q_-+k_1+k_2)^2-m^2,
\\ \nonumber
&& C_i=(k_i-p_+)^2-m^2, \qquad C=(k_1+k_2-p_+)^2-m^2,
\\ \nonumber
&& D_i=(q_++k_i)^2-m^2, \qquad D=(q_++k_1+k_2)^2-m^2.
\end{eqnarray}
For further integration it is useful to rewrite the denominators
in terms of the photon energy fractions $x_{1,2}$ and their emission
angles. In the case of the emission of both the photons along $p_-$ we
would have
\begin{eqnarray}
&& \frac{A}{m^2}=-x_1(1+z_1)
-x_2(1+z_2)+x_1x_2(z_1+z_2)+2x_1x_2\sqrt{z_1z_2}\cos\phi,
\nonumber \\
&& \frac{A_i}{m^2}=-x_i(1+z_i),
\end{eqnarray}
where $z_i=(\varepsilon\theta_i/m)^2$, $\phi$ is the azimuthal
angle between the planes containing the space vector pairs
$(\vecc{p}_-\, ,\vecc{k}_1)$ and $(\vecc{p}_-\, ,\vecc{k}_2)$.
In the same way one can obtain in the case $k_1\, ,k_2 \Vert q_-\,$:
\begin{eqnarray}
&& \frac{B}{m^2}=\frac{x_1}{y_1}(1+y_1^2z_1)
+\frac{x_2}{y_1}(1+y_1^2z_2)+x_1x_2(z_1+z_2)+2x_1x_2\sqrt{z_1z_2}\cos\phi,
\nonumber \\ &&
\frac{B_i}{m^2}=\frac{x_i}{y_1}(1+y_1^2z_i).
\end{eqnarray}
Then we perform the elementary azimuthal angle integration
and the integration over $z_1\, , z_2$ within the logarithmical
accuracy using the procedure suggested in paper~\cite{npm}:
\begin{eqnarray} \label{eq26}
\overline{a}=m^4\int\limits_{0}^{z_0}\d z_1 \int\limits_{0}^{z_0}\d z_2
\int\limits_{0}^{2\pi} \frac{\d \phi}{2\pi} a.
\end{eqnarray}
The list of the relevant integrals is given in Appendix A.
In this way one obtains the differential cross section in the collinear
region:
\begin{eqnarray} \label{sigc}
&& \d \sigma_{\mathrm{coll}} = \frac{\alpha^4 L}{4\pi^2 s}
\frac{\d^3q_+\d^3q_-}{q_+^0q_-^0}
\frac{\d x_1 \d x_2}{x_1x_2} \bigl(1+{\cal P}_{1,2} \bigr) \Biggl\{
\frac{1}{yr_1^2}
\biggl[\frac{1}{2}(L+2l)g_1g_5
\\ \nonumber && \quad
+(y^2+r_1^4)\ln\frac{x_2r_1^2}{x_1y}+
x_1x_2(y-x_1x_2)-2r_1g_5\biggr][B_{p_-p_-}\delta_{p_-p_-}
+ B_{p_+p_+}\delta_{p_+p_+}] \\ \nonumber && \quad
+\frac{1}{yr_1^2} \biggl[ \frac{1}{2}(L+2l+
4\ln y)g_1g_5+(y^2+r_1^4)\ln\frac{x_1r_1^2}{x_2y}
+x_1x_2(y-x_1x_2)-2r_1g_1\biggr] \\ \nonumber && \quad
\times [B_{q_-q_-}\delta_{q_-q_-}+B_{q_+q_+}\delta_{q_+q_+}]
+B_{p_-p_+}\delta_{p_-p_+}\biggl[(L+2l)\frac{g_1g_2}{r_1r_2}-2\frac{g_1}{r_1}
- 2\frac{g_2}{r_2}\biggr] \\ \nonumber && \quad
+B_{q_-q_+}\delta_{q_-q_+}\biggl[(L+2l
+2\ln(r_1r_2))\frac{g_1g_2}{r_1r_2}-2\frac{g_1}{r_1}-2\frac{g_2}{r_2}\biggr]
\\ \nonumber && \quad
+[B_{p_-q_-}\delta_{p_-q_-}+B_{p_+q_-}\delta_{p_+q_-}]
\biggl[(L+2l+2\ln y_1)\frac{g_1g_3}{r_1y_1t_1}-2\frac{g_1}{r_1}
-2\frac{g_3}{y_1t_1}\biggr] \\ \nonumber && \quad
+ [B_{p_+q_+}\delta_{p_+q_+}
+B_{p_-q_+}\delta_{p_-q_+}]\biggl[(L+2l+2\ln y_2)\frac{g_1g_4}{r_1y_2t_2}-
2\frac{g_1}{r_1}-2\frac{g_4}{y_2t_2}\biggr] \Biggr\}.
\end{eqnarray}
We used the symbol ${\cal P}_{1,2}$ for the interchange operator
(${\cal P}_{1,2}f(x_1,x_2)=f(x_2,x_1)$ ).
We used the notation (see also eq.~(\ref{eq:20})):
\begin{eqnarray}
l=\ln\biggl(\frac{\theta_0^2}{4}\biggr),\qquad g_5=y^2+r_1^2,
\end{eqnarray}
where $\theta_0$ is the collinear parameter. Delta--function
$\delta_{p,q}$ corresponds to the specific
conservation law of the kinematical situation defined
by the pair $p,q$ (see Table~1):
$\delta_{p,q} = \delta^{(4)}(\eta_2p_+ + \eta_1p_- - \lambda_1q_-
- \lambda_2q_+)$. Besides, we imply that the
first photon is emitted along the momentum $p$; and the second,
along the momentum $q$ ($p,\, q = p_-,\, p_+,\, q_-,\, q_+)$.
These $\delta$--functions could be taken into account in the integration
as is made in the expression for $\d I(\eta,\lambda)$
(see eq.~(\ref{di})).
Finally, we define
\begin{eqnarray}
B_{p,q}=\biggl(\frac{\eta_2s}{\lambda_1t}
+\frac{\lambda_1t}{\eta_2s}+1 \biggr)^2, \qquad
t=(p_- - q_-)^2.
\end{eqnarray}

\section{Contribution of the semi--collinear region}

We will suggest for definiteness that the photon with momentum $k_2$ moves
inside a narrow cone along the momentum direction of one of
the charged particles, while the other photon moves in any direction
outside that cone along any charged particle.
This choice allows us to omit the statistical factor
$1/2!$. The quasireal electron method~\cite{quasi} may be used to obtain
the cross section:
\begin{eqnarray}
\d \sigma^{\mathrm{sc}}&=&\frac{\alpha^4}{32s\pi^4}\,
\frac{\d^3q_-\d^3q_+\d^3k_1}
{q_-^0q_+^0k_1^0} V \frac{\d^3k_2}{k_2^0} \biggl\{
\frac{{\cal K}_{p_-}}{p_-k_2} \delta_{p_-}R_{p_-}
\nonumber \\  \label{eq30}
&+& \frac{{\cal K}_{p_+}}{p_+k_2} \delta_{p_+}R_{p_+}
+ \frac{{\cal K}_{q_-}}{q_-k_2} \delta_{q_-}R_{q_-}
+ \frac{{\cal K}_{q_+}}{q_+k_2} \delta_{q_+}R_{q_+} \biggr\}.
\end{eqnarray}
We omitted the terms of the kind $m^2/(p_-k_2)^2$ in eq. (\ref{eq30})
because their contribution does not contain the large logarithm $L$.
The quantities entering into eq.~(\ref{eq30}) are given by:
\begin{eqnarray}
V&=&\frac{s}{k_1p_+ \cdot k_1p_-} + \frac{s'}{k_1q_+ \cdot k_1q_-}
- \frac{t'}{k_1p_+ \cdot k_1q_+} - \frac{t}{k_1p_- \cdot k_1q_-}
\nonumber \\
&+& \frac{u'}{k_1p_+ \cdot k_1q_-} + \frac{u}{k_1q_+ \cdot k_1p_-}\, .
\end{eqnarray}
$V$ is the known accompanying radiation factor;
${\cal K}_i$ are the single photon emission collinear factors:
\begin{eqnarray}
{\cal K}_{p_-}={\cal K}_{p_+}=\frac{g_2}{x_2r_2}, \quad
{\cal K}_{q_-}=\frac{y_1^2+(y_1+x_2)^2}{x_2(y_1+x_2)}, \quad
{\cal K}_{q_+}=\frac{y_2^2+(y_2+x_2)^2}{x_2(y_2+x_2)}\, .
\end{eqnarray}
Quantities $R_i$ read:
\begin{eqnarray}
&& R_{p_-}=R[sr_2,tr_2,ur_2,s',t',u'], \nonumber
\quad R_{p_+}=R[sr_2,t,u,s',t'r_2,u'r_2], \\
&& R_{q_-}=R[s,t\frac{t_1}{y_1},u,s'\frac{t_1}{y_1},t',u'\frac{t_1}{y_1}],
\quad R_{q_+}=R[s,t,u\frac{t_2}{y_2},s'\frac{t_2}{y_2},t'\frac{t_2}{y_2},u'],
\end{eqnarray}
where the function $R$ has the form~\cite{ber81}:
\begin{eqnarray}
&& R[s,t,u,s',t',u']=\frac{1}{ss'tt'}\bigl[ ss'(s^2+{s'}^2)+tt'(t^2+{t'}^2)
+uu'(u^2+{u'}^2) \bigr], \nonumber \\ \nonumber
&& s=(p_++p_-)^2, \quad s'=(q_++q_-)^2, \quad t=(p_--q_-)^2, \\
&& t'=(p_+-q_+)^2, \quad u=(p_--q_+)^2, \quad u'=(p_+-q_-)^2.
\end{eqnarray}
Finally, we define
\begin{eqnarray}
\delta_{p_-}&=&\delta^{(4)}(p_-r_2+p_+-q_+-q_--k_1),\nonumber \\ \nonumber
\delta_{p_+}&=&\delta^{(4)}(p_-+p_+r_2-q_+-q_--k_1), \\ \nonumber
\delta_{q_-}&=&\delta^{(4)}(p_-+p_+-q_+-q_-\frac{y_1+x_2}{y_1}-k_1), \\
\delta_{q_+}&=&\delta^{(4)}(p_-+p_+-q_+\frac{y_2+x_2}{y_2}-q_--k_1).
\end{eqnarray}

Performing the integration over angular variables of the collinear photon
we obtain
\begin{eqnarray}
\d \sigma^{\mathrm{sc}} &=& \frac{\alpha^4 L}{16 s\pi^3}\,
\frac{\d^3q_-\d^3q_+\d^3k_1}
{q_-^0q_+^0k_1^0} \d x_2 V \biggl\{{\cal K}_{p_-} [R_{p_-}\delta_{p_-}
+ R_{p_+}\delta_{p_+}]         \nonumber \\ \label{eq35}
&+& \frac{1}{y_2}{\cal K}_{q_+} R_{q_+}\delta_{q_+}
+ \frac{1}{y_1}{\cal K}_{q_-} R_{q_-}\delta_{q_-} \biggr\}.
\end{eqnarray}

To see that the sum of cross sections
(\ref{sigc}) and (\ref{eq35})
\begin{eqnarray} \label{siggg}
\d \sigma^{\gamma\gamma}=\d \sigma^{\mathrm{coll}}
+ \int\d O_1 \bigl(\frac{\d \sigma^{\mathrm{sc}}}{\d O_1} \bigr)
\end{eqnarray}
does not depend on the auxiliary parameter $\theta_0$.
We verify that terms
$L \cdot l$ from eq.~(\ref{sigc}) cancel out with the terms
\begin{equation}
L \frac{k_1^0q_i^0}{2\pi} \int\frac{\d O_1}{k_1q_i} \ \approx \ - L\cdot l,
\end{equation}
which arise from 16 regions in the semi--collinear kinematics.

\section{Numerical results and discussion}

We separated the contribution of the collinear and semi--collinear
regions using the auxiliary parameter $\theta_0$. By direct numerical
integration according to the presented formulae we had convinced
ourselves that the total result is independent on the choice of $\theta_0$.

It is convenient to compare the cross section of double hard
photon emission with the Born cross section
\begin{eqnarray}
\sigma^{\mathrm{Born}}=\frac{\alpha^2\pi}{2s}
\int\limits_{-\cos\psi_0}^{\cos\psi_0}\left(\frac{3+c^2}{1-c}\right)^2\d c.
\end{eqnarray}
For illustrations we integrated over some typical experimental
angular acceptance and chose the following values of the parameters:
\begin{eqnarray}
&& \psi_0=\pi/4,\quad \sqrt{s}=0.9\; \mbox{GeV}, \quad \Delta_1=0.4, \quad
\Delta=0.05, \quad \theta_0 = 0.05, \nonumber \\
&& L = 15.0, \quad l=-7.38\, ,
\end{eqnarray}
where $\Delta_1$ defines the energy threshold for the registration
of the final electron and positron: $q_{\pm}^0 > \varepsilon_{\mathrm{th}}
=\varepsilon\Delta_1$.
Note that restrictions on $\theta_0$ (\ref{eq6}) and (\ref{z0})
$(z_0=\exp\{L+l\} \gg 1)$ are fulfilled.

For the chosen parameters we get
\begin{eqnarray}
&& \sigma^{\mathrm{Born}} = 1.2\; \mbox{mkb}, \qquad \nonumber
   \frac{\sigma^{\mathrm{coll}}}
        {\sigma^{\mathrm{Born}}}\cdot 100\% = - 0.25\,\%, \\
&& \frac{\sigma^{\mathrm{sc}}}
        {\sigma^{\mathrm{Born}}}\cdot 100\% =   0.81\,\%, \qquad
\delta\sigma^{\mathrm{tot}}=
  \frac{\sigma^{\mathrm{sc}}+\sigma^{\mathrm{coll}}}
        {\sigma^{\mathrm{Born}}}\cdot 100\% =   0.56\,\%.
\end{eqnarray}
The {\em phenomenon\/} of negative contribution to the cross section
from the collinear kinematics is an artifact of our approach.
Namely, we systematically omitted positive terms without large logarithms,
among them we dropped terms proportional to $l^2$. The cancellation
of $l^2$ terms can be seen only after adding the contribution
of the non--collinear kinematics (when both photons are emitted
outside narrow cones along charged--particle momenta). The non--collinear
kinematics does not provide any large logarithm $L$.

Both quantities $\sigma^{\mathrm{coll}}$ and
$\sigma^{\mathrm{sc}}$ depend on auxiliary parameter
$\theta_0$. We eliminated by hands from eq.~(\ref{sigc}) the
terms proportional to $l$ and obtained the following quantity:
\begin{eqnarray}
   \frac{\sigma_{\mathrm{coll}}^{\mathrm{bare}}}
       {\sigma^{\mathrm{Born}}}\cdot 100\% = 1.43\,\%.
\end{eqnarray}
This quantity corresponds to an approximation for the correction
under consideration in which one considers only the collinear
regions and takes into account only terms proportional to $L^2$ and $L$
(all terms dependent on $\theta_0$ are to be omitted).
Having in mind the cancellation of $\theta_0$--dependence in
the sum of the collinear and semi--collinear contributions,
we may subtract from the value of the semi--collinear contribution
the part which is associated with $l$:
\begin{eqnarray}
&& \sigma_{\mathrm{sc}}^{\mathrm{bare}}= \nonumber
\sigma_{\mathrm{sc}} + (\sigma_{\mathrm{coll}}
- \sigma_{\mathrm{coll}}^{\mathrm{bare}} ), \qquad
\frac{\sigma_{\mathrm{sc}}^{\mathrm{bare}}}
       {\sigma^{\mathrm{Born}}}\cdot 100\% = -0.87\,\%.
\end{eqnarray}
Looking at {\em bare\/} quantities one can get an idea of relative
impact of two considered regions. We see that at the precision
level of $0.1\%$ the next--to--leading contributions of
semi--collinear regions are important.

In figure~1 we illustrated the dependence on parameter $\Delta$
of the bare collinear contribution for different fixed values
of $\Delta_1$. Large growing in the region of small $\Delta$
corresponds to an infrared singularity, which will be cancelled
after adding contributions of virtual and soft photon
emission.

\ack

We are grateful to S.~Eidelman, G.~Fedotovich, P.~Franzini and
G.~Pancheri for fruitful discussions. For the help in numerical
calculations we thank I.V.~Amirkhanov, T.A.~Strizh and T.P.~Puzynina.
We are also grateful to INFN, Parma university for hospitality.
The work was partially supported by INTAS grant 93--1867
and RFBR grant 96-02-17512.
One of us (A.B.A.) is thankful to the INTAS fundation
for financial support via an ICFPM grant.

\section*{Appendix A}

We present here the list of integrals (see eqs.~(\ref{abcd} -- \ref{eq26})):
\begin{eqnarray}
&& \overline{\frac{A_2}{A^2A_1}}=\frac{L_0}{x_1x_2r_1^2}
\biggl[\frac{1}{2}L_0 + \ln \frac{x_2r_1^2}{x_1y} - 1
+ \frac{x_1x_2}{y} \biggr], \nonumber \\
&& \overline{\frac{1}{AA_1}}=\frac{L_0}{x_1x_2r_1}
\biggl[\frac{1}{2}L_0 + \ln \frac{x_2r_1^2}{x_1y} \biggr],
\qquad \overline{\frac{m^2}{AA_1^2}}=-\frac{L_0}{x_1^2x_2r_1}\, ,
\\ \nonumber
&& \overline{\frac{1}{A_1A_2}}=\frac{L_0^2}{x_1x_2},
\qquad \overline{\frac{1}{A_1B_2}}=-\frac{L_0}{y_1x_1x_2}
(L_0+2\ln y_1),
\\ \nonumber
&& L_0=\ln z_0 \equiv L + l, \qquad l=\ln(\frac{\theta_0^2}{4}),
\qquad L=\ln(\frac{4\varepsilon^2}{m^2}).
\end{eqnarray}
The remaining integrals could be obtained by simple substitutions
defined in eqs.~(\ref{abcd} -- \ref{eq26}).

\newpage

\begin{figure}[l]
\begin{center}
\epsfbox[0 0 300 300]{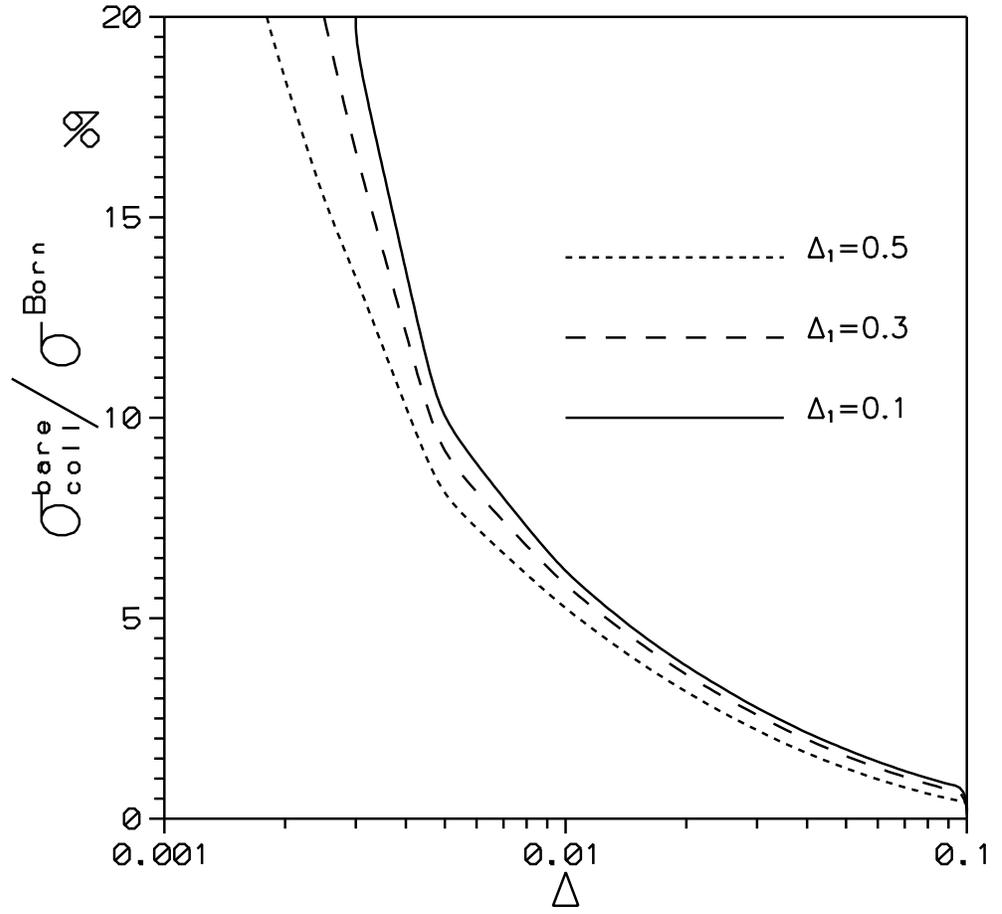}
\end{center}
\caption{
The ratio {$\sigma_{\mathrm{coll}}^{\mathrm{bare}}/
\sigma^{\mathrm{Born}}$} in percent
as functions of {$\Delta$}
for {$\psi_0=\pi/4$} and different values of {$\Delta_1$}.
}
\label{Fig1}
\end{figure}

\end{document}